\documentstyle[twoside,fleqn,espcrc2,epsf]{article}

\newcommand{\ppri}{p^\prime}
\newcommand{\pprimu}{p^{\prime\mu}}

\newcommand{\vev}[1]{\langle #1 \rangle}

\newcommand{\act}{{\cal S}}

\newcommand{\nl}{\nonumber \\}

\newcommand{\be}{\begin{equation}}
\newcommand{\ee}{\end{equation}}

\newcommand{\AmS}{{\protect\the\textfont2
  A\kern-.1667em\lower.5ex\hbox{M}\kern-.125emS}}

\title{ Can One Study Heavy Meson Semileptonic Decays on Coarse 
Anisotropic Lattices ?}

\author{
        Junko Shigemitsu
        \address{Physics Department, 
        The Ohio State University, 
        Columbus, OH 43210, USA.}
        \thanks{Talk presented 
at LATTICE 2000, Bangalore, India, August 2000.
Describes work in collaboration with S.Collins, C.T.H.Davies, J.Hein, 
R.R.Horgan and G.P.Lepage. }
                           }
       
\begin{document}

\begin{abstract}

Exploratory studies of heavy meson semileptonic decays on coarse 
lattices using Symanzik improved glue, NRQCD heavy and D234 light quarks 
are discussed.  Comparisons are made between calculations on 
anisotropic and isotropic lattices.  We find evidence that having an 
anisotropy helps in extracting better signals at higher momenta.

\end{abstract}

\maketitle

\section{Introduction}
Semileptonic decays of heavy mesons into light hadrons such as, 
$B \rightarrow \pi \,  l \nu$ or
$B \rightarrow \rho \, l \nu$, involve hadrons with sizeable 
momenta, once one moves away from the zero-recoil point at 
$q^2 \equiv (p - p^\prime)^2 = q^2_{max}$. ``$p_\mu$'' is the momentum 
of the decaying heavy meson and ``$p^\prime_\mu$'' that of the light 
daughter meson.  High momenta introduce both noise and $ap$ discretization 
errors into lattice simulations making calculations of semileptonic 
form factors that extend 
over a wide range in $q^2$ a challenging task.  We have carried 
out exploratory simulations to investigate whether anisotropic 
lattices can help 
with the signal-to-noise problem.  We study two- and three-point correlators 
involving finite momentum hadrons on both anisotropic and isotropic lattices
 with comparable coarse spatial lattice spacings.  Working with identical 
source operators and smearings on the initial time slices, we compare 
the quality of signals that can be extracted at later times.  We find 
that it is considerably easier to obtain good signals on anisotropic 
lattices at higher momenta (e.g. once momenta reach 
(1,1,1)$2 \pi/(a_s L)$ and beyond).

\section{Some Calculational Details}

The gauge actions on the isotropic and anisotropic lattices 
correspond to Symanzik improved actions and include both plaquettes 
$P_{\mu \nu}$    and rectangles $R_{\mu \nu}$ \cite{weisz}.
\begin{eqnarray}
\label{isoglue}
& &\act^{(iso)}_G = \nl
&& - \beta \sum_{x,\,\mu > \nu}  \left\{
\frac{5}{3} \frac{P_{\mu \nu}}{u_L^4} 
- \frac{1}{12} \frac{R_{\mu \nu}}{u_L^6} 
- \frac{1}{12} \frac{R_{\nu \mu}}{u_L^6} \right\}   
\end{eqnarray}

\begin{eqnarray}
& &\act^{(aniso)}_G  =\nl
 && - \beta
 \sum_{x,\,s > s^\prime} \frac{1}{\chi_0} \left\{
\frac{5}{3} \frac{P_{ss^\prime}}{u_s^4} 
- \frac{1}{12} \frac{R_{ss^\prime}}{u_s^6} 
- \frac{1}{12} \frac{R_{s^\prime s}}{u_s^6} \right\} \nl
 & & - \beta \;\; \sum_{x,s} \chi_0 \left\{
\frac{4}{3} \frac{P_{st}}{u_s^2 u_t^2} 
- \frac{1}{12} \frac{R_{st}}{u_s^4u_t^2} \right\} .
\label{anisoglue}
\end{eqnarray}
We use the Landau link definition for the tadpole improvement 
factors $u_L$, $u_s$ and $u_t$.  $\chi_0$ is the bare anisotropy. It is 
related to the renormalized anisotropy $\chi = a_s/a_t$ as,
$\chi_0 =  \chi/\eta$.
$\eta$ must be fixed using some physics criterion, such as the requirement 
of a correct relativistic dispersion relation for 
torelons \cite{ron}. For the light quarks we use the D234 action \cite{alford},
\begin{eqnarray}
\label{sd234c}
&& \act_{D234} = 
  \sum_x \overline{\Psi}
\left\{  \gamma_t  (\nabla_t - \frac{1}{6} 
C_{3t} \nabla_t^{(3)}) \right.  \nl
&&  + \frac{C_0}{\chi} \vec{\gamma} \cdot (\vec{\nabla} -
 \frac{1}{6} 
C_{3} \vec\nabla^{(3)}) 
 + a_t m_0  \nl
 &  & - \frac{r }{2} \left [ \chi ( \nabla_t^{(2)} 
- \frac{1}{12} C_{4t} \nabla_t^{(4)} )
 +  \frac{1}{\chi} \sum_j ( \nabla_j^{(2)} \right.  \nl
&& \left. \left. - \frac{1}{12} C_{4} \nabla_j^{(4)} ) \right ] 
 - r  \frac{C_F}{4} 
i \sigma_{\mu \nu} \tilde{F}^{\mu \nu} \frac{a_s a_t}{a_\mu a_\nu} 
\right\} 
\Psi .
\end{eqnarray}
For the isotropic simulations $\chi$ and all the $C_j$ coefficients 
are set equal to one.  On anisotropic lattices we drop discretization 
corrections in the temporal direction and $C_{3t} = C_{4t}= 0$. The 
coefficient $C_0$ must be tuned to ensure correct dispersion relations 
in correlators involving light propagators.  We use a one-loop 
perturbative estimate for $C_0$ in our anisotropic simulations \cite{pert}. 
The other coeffcients $C_3$, $C_4$ and $C_F$ are set equal to one. 
 For 
the heavy quarks we employ the NRQCD action \cite{cornell}, 
suitably modified to 
allow for an anisotropy. 

\vspace{.1in}
\noindent
In Table I we summarize simulation parameters.  In this exploratory 
study we have not attempted to tune quark masses very carefully.  We work 
at one light quark mass slightly heavier than the strange quark.  
The heavy quark mass is chosen so that the heavy-light pseudoscalar meson 
mass is close to that of the physical $B_s$.  On the anisotropic 
lattice we also worked at a second heavy quark mass value close to 
the charm.
\begin{table}[t]
\caption{Simulation Details.  
  }
\begin{center}
\begin{tabular}{c|cc}
\hline
    & isotropic& anisotropic\\\hline
 lattice size  &  $8^3 \times 20$    &  $ 8^3 \times 48 $  \\
\# configs     &  200                &   200              \\
$\beta$        &  1.719              &  1.8               \\
$\chi_0$       &        1            &  6.0             \\
$\chi = a_s /a_t$ &     1            &  5.3 \cite{ron}      \\
$C_0$          &        1            &    0.82 \cite{pert}        \\
$a_s^{-1}$     &        0.8(1) GeV \cite{alford}   &   0.7(1) GeV \cite{ron} \\
$a_t^{-1}$    &         0.8(1) GeV $\quad$  &   3.7(4) GeV $\quad$   \\
$a_t m_0$     &         1.15         &   0.39          \\
$P/V$         &         0.725(5)     &   0.726(6)     \\
$a_sM_0$      &         6.5          &  7.0 and 2.0    \\
\end{tabular}
\end{center}
\end{table}

\section{Two Point Correlators}
\begin{figure}
\epsfxsize=7.0cm
\centerline{\epsfbox{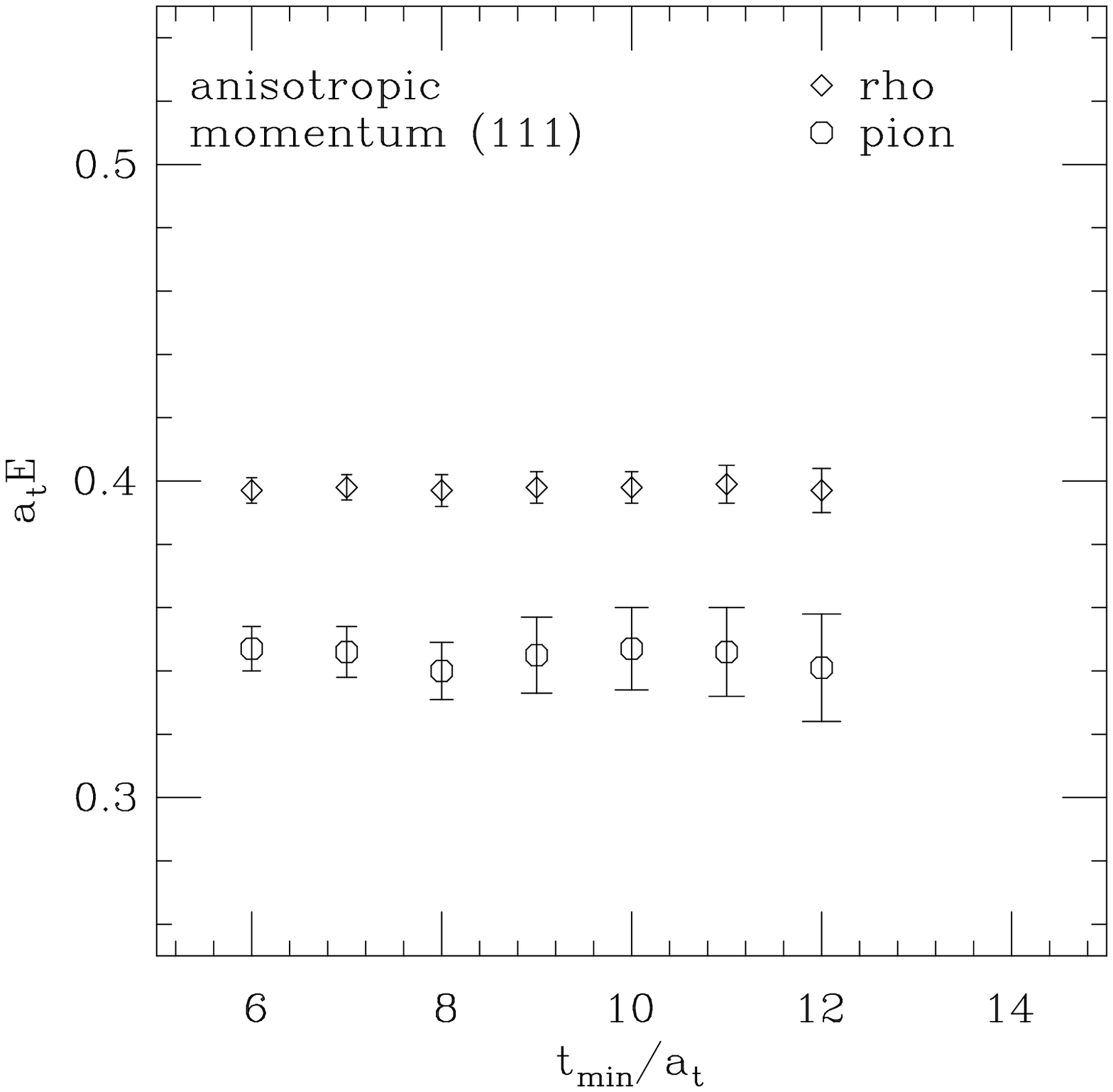}}
\end{figure}

\begin{figure}
\epsfxsize=7.0cm
\centerline{\epsfbox{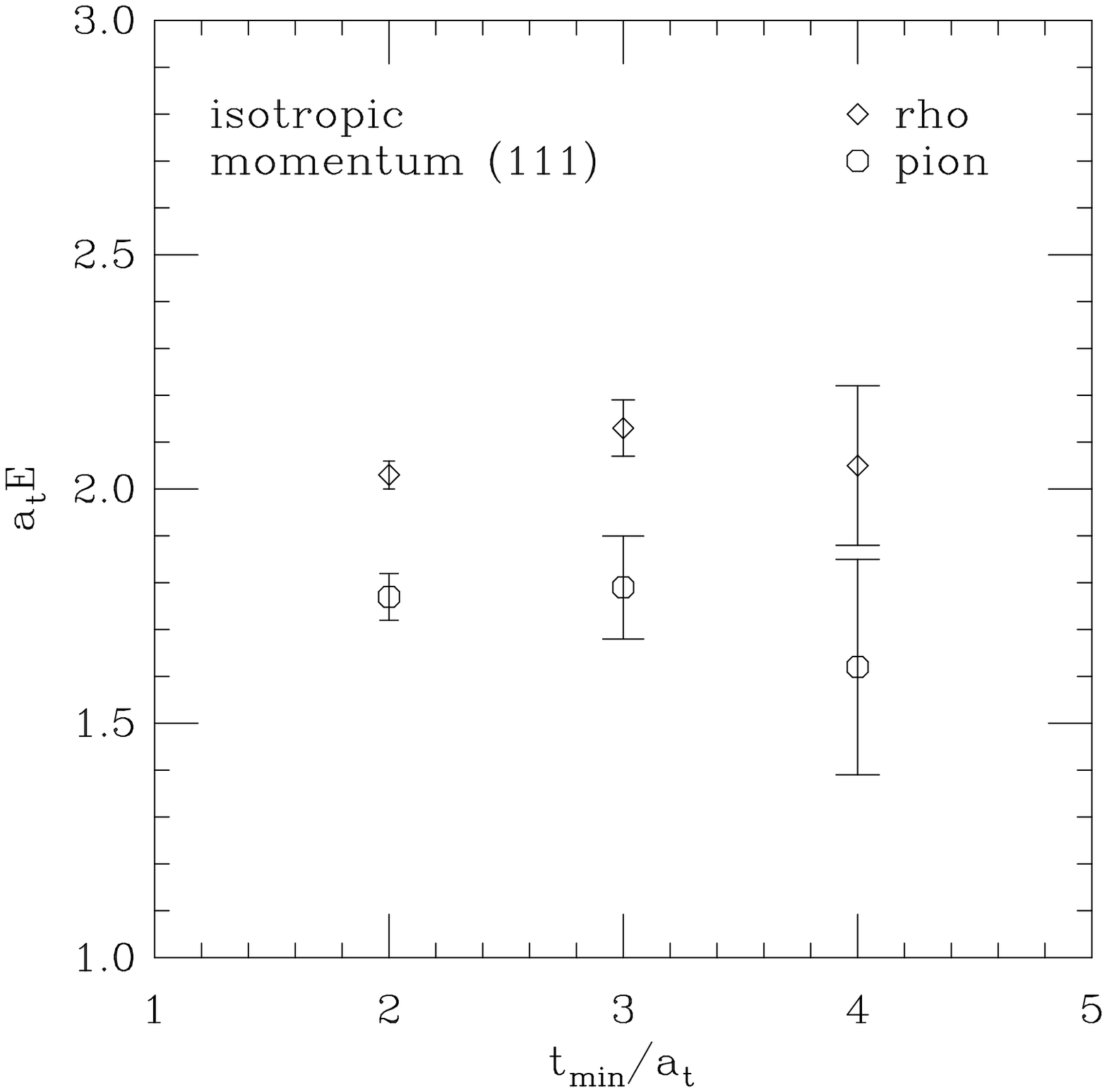}}
\caption{
Fitted energies versus $t_{min}/a_t$ for the pion and the rho for 
momentum (1,1,1) on anisotropic (top figure) and 
isotropic (bottom figure) lattices.
  }
\end{figure}

\begin{figure}
\epsfxsize=7.0cm
\centerline{
\epsfbox{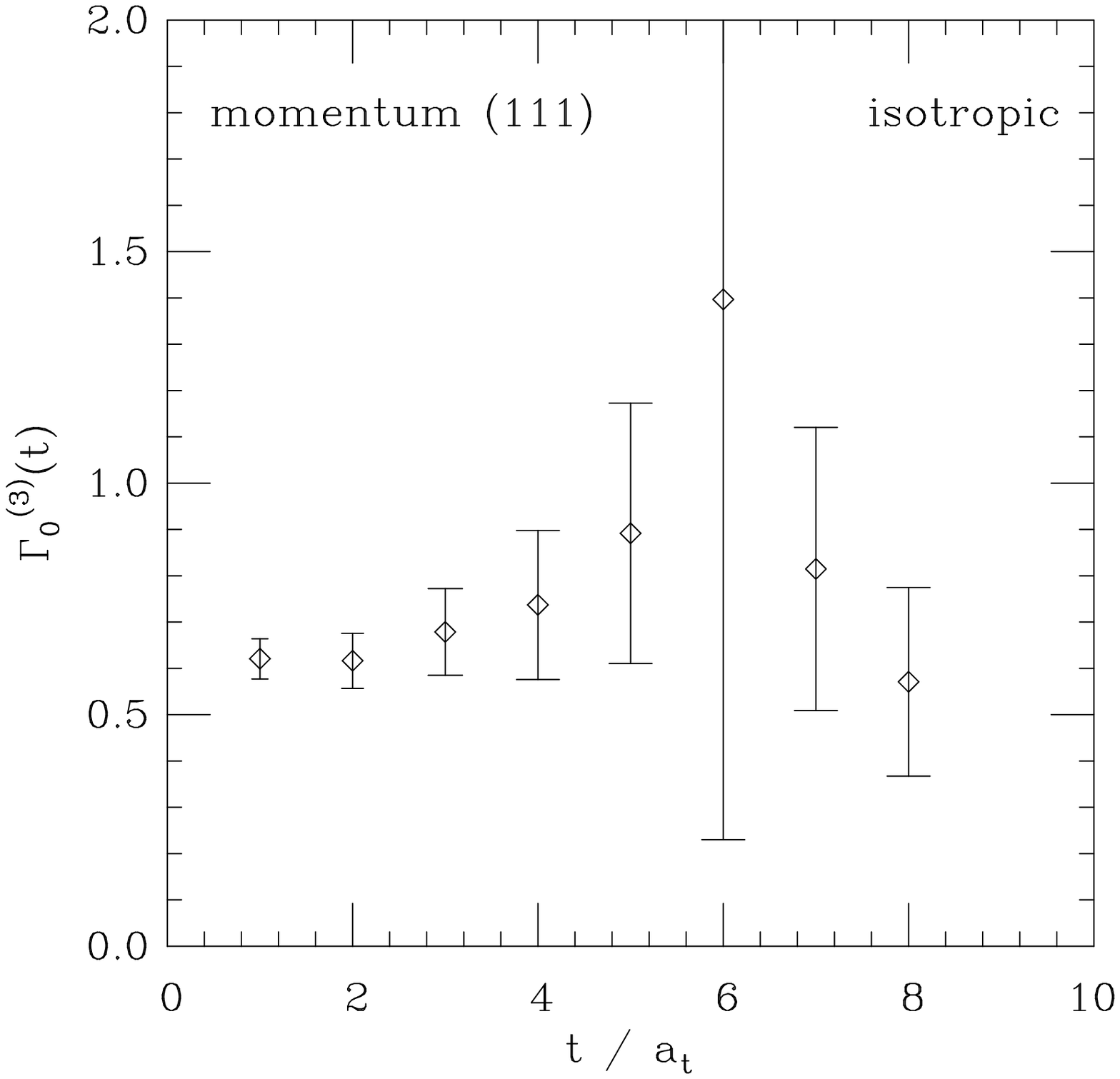}
}
\caption{ $\Gamma^{(3)}_0(t)$ of eq.(5) for pion momentum  (1,1,1)
from isotropic lattices versus time in lattice units.
}
\end{figure}
\begin{figure}
\epsfxsize=7.0cm
\centerline{
\epsfbox{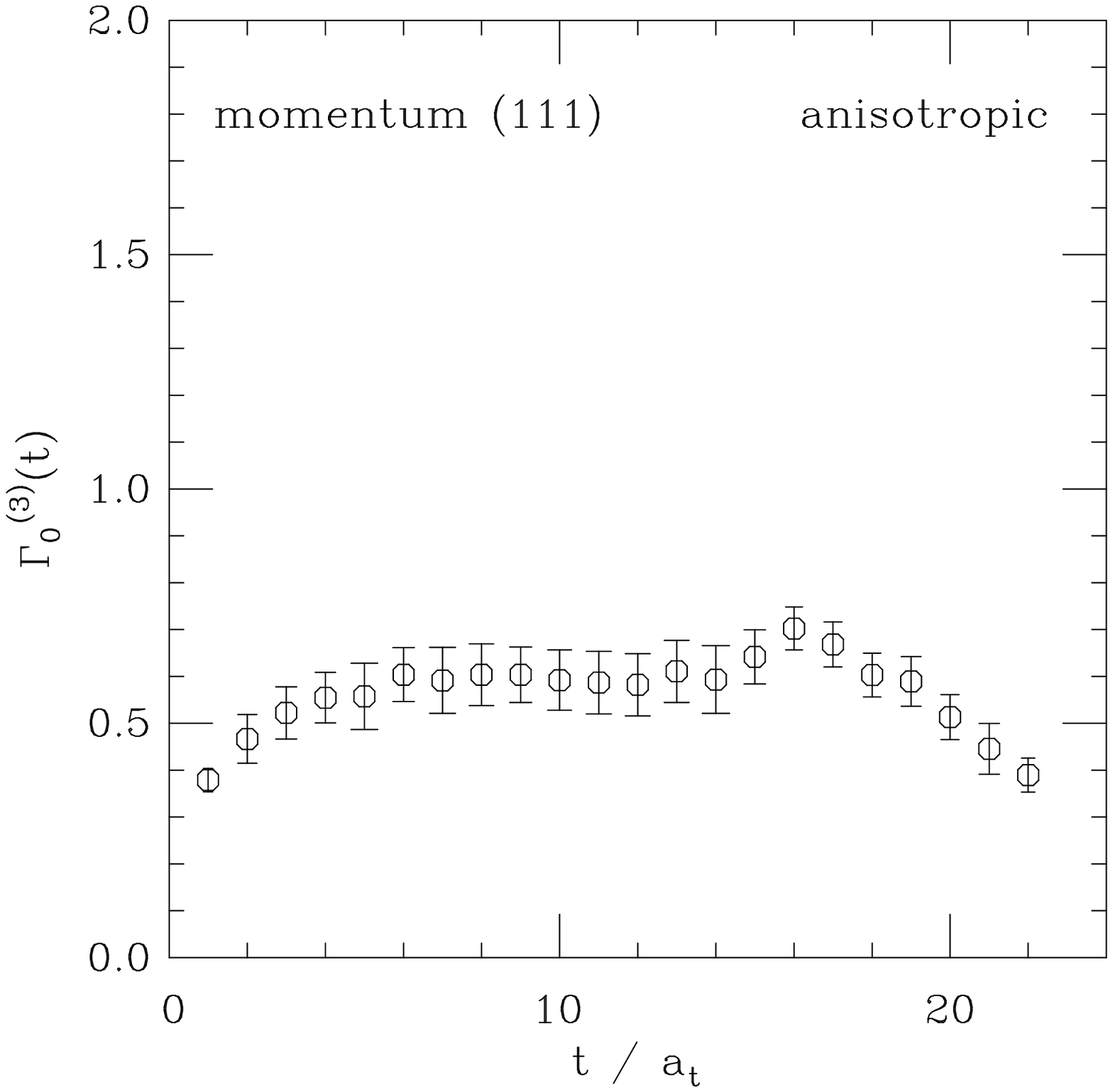}
}
\caption{ Same as Fig. 2. from anisotropic lattices.
}
\end{figure}

\begin{figure}
\epsfxsize=7.0cm
\centerline{
\epsfbox{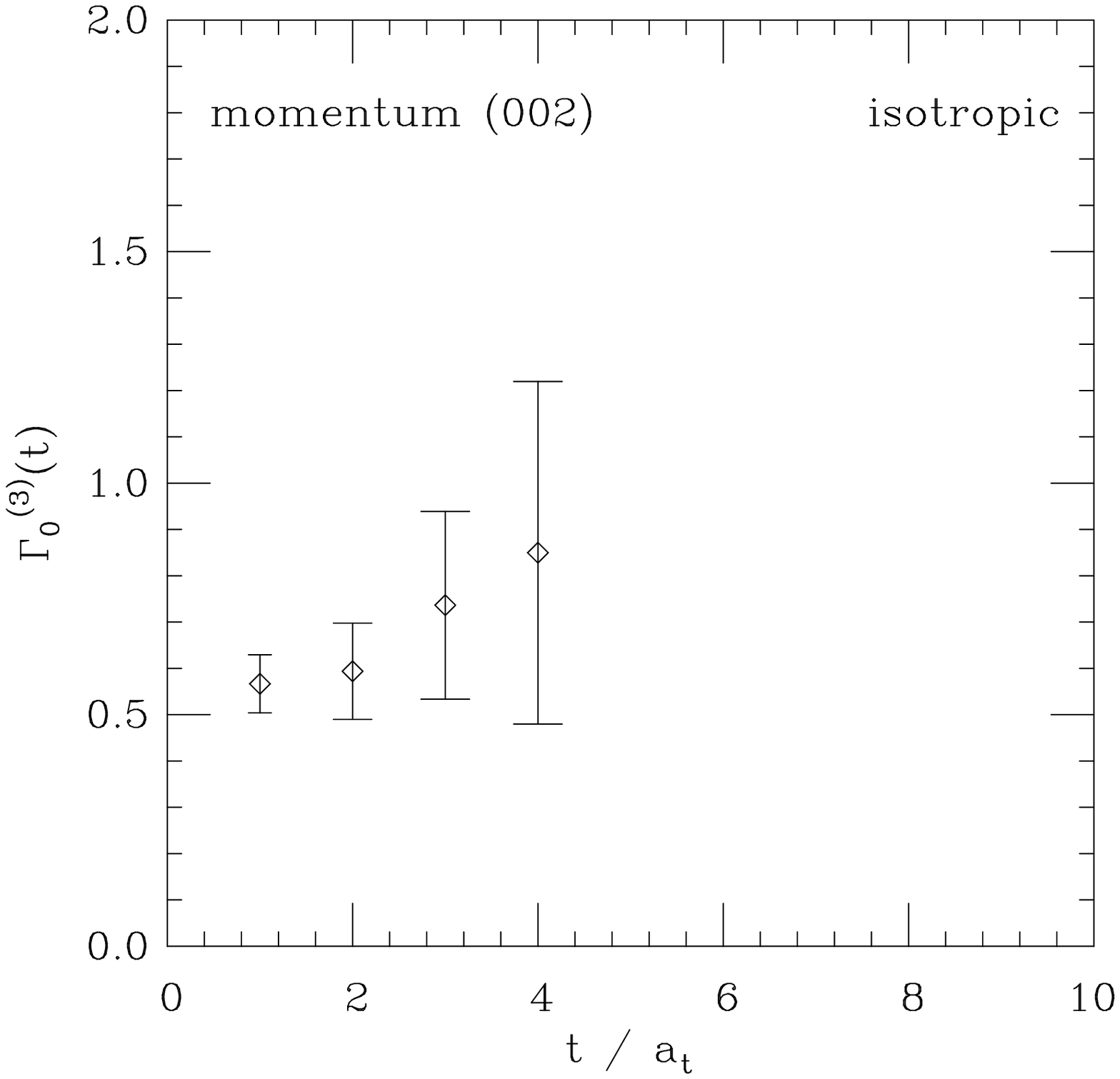}
}
\caption{ $\Gamma^{(3)}_0(t)$ of eq.(5) for pion momentum  (0,0,2)
from isotropic lattices versus time in lattice units.
}
\end{figure}

\begin{figure}
\epsfxsize=7.0cm
\centerline{
\epsfbox{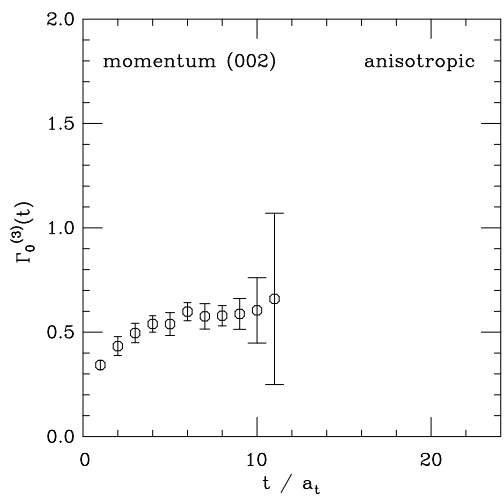}
}
\caption{ Same as Fig. 4. from anisotropic lattices.
}
\end{figure}

In Figure 1. we show examples of extracted finite momentum  energies from 
the pion and rho channels at momentum (1,1,1)$2 \pi/(a_sL)$.  
Since identical signals are being created on the initial time slices,
 effective mass plateaus are of the same length in physical units 
on the isotropic and anisotropic lattices.
  However, on the anisotropic 
lattice the region where the signal is still 
appreciably above the noise is being probed much more 
frequently.  It is then easier to recognize the onset of a plateau and 
 extract fitted energies with confidence.

\section{Three Point Correlators}
Semi-leptonic form factors for pseudoscalar to pseudoscalar 
decays are defined through the hadronic matrix element of 
the heavy-light vector current $V_\mu$,
\begin{eqnarray}
\label{vmu}
\vev{\pi(\ppri)|V^\mu|B(p)} &=& f_+(q^2) (p^\mu + \pprimu)  +\nl 
                             && f_-(q^2) (p^\mu - \pprimu) .
\end{eqnarray}
In lattice simulations the above matrix element can be obtained from 
ratios of three- and two-point correlators.  We define, 
\begin{eqnarray}
\label{c3}
&&\Gamma^{(3)}_\mu(\vec{p},\vec{\ppri},t_B,t)  \equiv  \nl
&& \frac{G^{(3)}_\mu(\vec{p},\vec{\ppri},t_B,t)}{G_B^{(2)}(\vec{p},t_B-t) \,
G^{(2)}_\pi(\vec{\ppri},t)} \sqrt{\zeta_{BB} \zeta_{\pi\pi}} \\
\label{vtilde}
&&\rightarrow  \frac{\vev{B(\vec{p})|V^\mu|\pi(\vec{\ppri})}}
{2\sqrt{E_B E_\pi}} ,
\end{eqnarray}
where,
\begin{eqnarray}
\label{threepnt1}
&&G_\mu^{(3)}(\vec{p},\vec{\ppri},t_B,t) = \sum_{\vec{x}}\sum_{\vec{y}}
e^{-i\vec{p}\cdot\vec{x}} e^{i(\vec{p} - \vec{\ppri})\cdot \vec{y} } \nl 
&& \quad \quad \vev{0|\Phi_{B}(t_B,\vec{x})\,V^L_\mu(t,\vec{y})\, \Phi^\dagger_\pi(0)
|0} 
\end{eqnarray}
and
\begin{eqnarray}
&&G^{(2)}_B(\vec{p},t) = \nl
&& \sum_{\vec{x}}e^{-i\vec{p}\cdot\vec{x}}
\vev{0|\Phi_B(t,\vec{x})\,\Phi_B^\dagger(0)|0}  \nl
&& \; \qquad \qquad \rightarrow 
 \zeta_{BB}e^{-E^{sim}_Bt} ,  \\
\label{pioncorr}
&&G^{(2)}_\pi(\vec{\ppri},t) \rightarrow
 \zeta_{\pi\pi}\,e^{-E_\pi t} .
\end{eqnarray}
$\Phi_B$ and $\Phi_\pi$ are interpolating operators for the $B$ and $\pi$ 
mesons.  $V^L_\mu$ is the lattice heavy-light vector current
 appropriately matched to the continuum. Figures 2. - 5. show 
$\Gamma^{(3)}_{\mu=0}$ 
versus $t/a_t$ at fixed $t_B$ for two different pion momenta 
$\vec{\ppri}$ on the two lattices.  The B meson momentum is kept at 
$\vec{p} = 0$.  
One sees again that the anisotropy greatly facilitates 
extracting a reliable signal.  From this data one can use eq.(\ref{vmu}) 
and eq.(\ref{vtilde}) to determine the form factors $f_\pm(q^2)$.  
Results for form factors will be presented elsewhere
\cite{aniso}.

\section{Summary}
We find evidence that anisotropic simulations allow us to go 
to higher momenta in two- and three-point hadronic correlators. 
For studies of semi-leptonic heavy meson decays this means that 
a wider range in $q^2 = (p-\ppri)^2$ can be covered.
As part of the present project we have also tested for $ap$ discretization 
errors in our simulations.  We have looked at dispersion relations and 
the effects of nonzero B meson momenta on the decay constant $f_B$.  
We find good continuum behavior up to $a_sp \approx 1.5$ and 
only 5-10\% deviations up to about $a_sp \sim 2$.
These tests will also be summarized in a forthcoming paper together 
with the form factor results \cite{aniso}.  
There we will go into details of 
weighing the additional costs associated with anisotropic simulations 
against the advantages described here.

\vspace{.1in}
\noindent
Acknowledgements : This work was supported by DOE Grant
 DE-FG02-91ER40690.  The numerical simulations were carried out 
at the Ohio Supercomputer Center and at NERSC.

\end{document}